# AI vs. Human Judgment of Content Moderation: LLM-as-a-Judge and Ethics-Based Response Refusals

Stefan Pasch[1]

## Abstract

As large language models (LLMs) are increasingly deployed in high-stakes settings, their ability to refuse ethically sensitive prompts—such as those involving hate speech or illegal activities—has become central to content moderation and responsible AI practices. While refusal responses can be viewed as evidence of ethical alignment and safety-conscious behavior, recent research suggests that users may perceive them negatively. At the same time, automated assessments of model outputs are playing a growing role in both evaluation and training. In particular, LLM-as-a-Judge frameworks—in which one model is used to evaluate the output of another—are now widely adopted to guide benchmarking and fine-tuning. This paper examines whether such model-based evaluators assess refusal responses differently than human users. Drawing on data from Chatbot Arena and judgments from two AI judges (GPT-4o and Llama 3 70B), we compare how different types of refusals are rated. We distinguish ethical refusals, which explicitly cite safety or normative concerns (e.g., "I can't help with that because it may be harmful"), and technical refusals, which reflect system limitations (e.g., "I can't answer because I lack real-time data"). We find that LLM-as-a-Judge systems evaluate ethical refusals significantly more favorably than human users, a divergence not observed for technical refusals. We refer to this divergence as a *moderation bias*—a systematic tendency for model-based evaluators to reward refusal behaviors more than human users do. This raises broader questions about transparency, value alignment, and the normative assumptions embedded in automated evaluation systems.

## 1. Introduction

As large language models (LLMs) are deployed at scale across increasingly sensitive domains, concerns about their ethical alignment, safety, and trustworthiness have emerged as key priorities for developers, researchers, and policymakers (Hagendorff, 2020). A core strategy for addressing these concerns is for models to refuse to answer certain prompts—particularly those involving illegal, harmful, or morally controversial content. For instance, when asked how to engage in criminal activity or hate speech, an aligned model might respond: *"I'm sorry, I can't assist with that request because it involves unsafe content."* These response refusals now play a central role in content moderation and responsible AI development (Xie et al., 2024). Yet their reception remains contested: while refusals can be seen as evidence of model safety and normative compliance, recent work by Pasch (2025) has shown that users frequently rate refusal responses negatively—potentially because they are experienced as evasive, overly moralizing, or uncooperative.

In parallel to growing concerns around model alignment, LLMs are also increasingly used to evaluate one another. A prominent approach is the LLM-as-a-Judge (LaaJ) framework, in which one model assesses and ranks the outputs of other models (Zheng et al., 2023). While LaaJ methods are praised for their scalability and consistency with human raters, recent research has also revealed systematic biases—for example, a tendency to prefer responses that are longer, more emotionally neutral, or confidently phrased (Ye et al., 2024). Yet little is known about how these frameworks evaluate content moderation behaviors, particularly response refusals.

---

[1] Division of Social Science & AI, Hankuk University of Foreign Studies: stefan.pasch@outlook.com

Studying how LaaJ systems respond to refusals is especially important given their expanding role in shaping the overall behavior of AI systems, including how they navigate ethically sensitive situations. LaaJ frameworks are no longer used solely for evaluation after deployment; they are increasingly integrated into processes of training, benchmarking, and model selection—thus helping to define what kinds of outputs are considered acceptable, safe, or desirable (Wu et al., 2024; Saha et al., 2024). This influence shapes how content moderation is implemented by LLMs at scale—raising important concerns about transparency, accountability, and public legitimacy in AI governance (Mökander & Floridi, 2021; Whittlestone et al., 2021). In fact, since leading AI developers have emphasized the importance of aligned and safety-conscious model behavior (OpenAI, 2024; Meta 2024a; Anthropic, 2023), LaaJ frameworks may interpret refusal responses as signs of responsible alignment—whereas human users might perceive them as overly cautious, evasive, or uncooperative. To investigate this potential divergence in how refusal responses are judged, we pose the following research question:

**RQ:** *Do LLM-as-a-Judge systems evaluate refusal responses—particularly those grounded in ethical alignment—differently than human users do?*

To address this question, we draw on data from Chatbot Arena (Chiang et al., 2024)—a large-scale platform where human users evaluate paired LLM responses to the same prompt—and compare these user preferences to judgments from two state-of-the-art LLM-as-a-Judge systems: GPT-4o and Llama 3 70B. Using this setup, we analyze how different forms of refusal behavior affect win/loss outcomes across human and model-based evaluations. In particular, we distinguish between ethical refusals, where the model declines a prompt due to normative concerns (e.g., legality, harm, or moral risk), and technical refusals, where the refusal stems from system limitations (e.g., lack of real-time data or model capability). We further separate full refusals from disclaimers, in which the model qualifies its response without outright rejecting the prompt.

Our results reveal a consistent divergence: LaaJ systems evaluate ethical refusals more favorably than human users, a difference not observed for technical refusals or disclaimers. These findings highlight the presence of what we term a *moderation bias* in LaaJ evaluation and raise broader questions about transparency, value alignment, and contestability in automated model assessment.

Notably, this paper does not make normative claims about the "correct" level of refusals or whether users or model-based evaluators are ultimately right in their assessments. Rather, our aim is to document and explain a systematic gap in how refusal responses—particularly those driven by ethical alignment—are evaluated by human users versus LaaJ systems.

By doing so, the paper contributes to three ongoing conversations:

1. It extends empirical research on model alignment and safety by examining how aligned behaviors are evaluated—not just produced—by LLMs.
2. It introduces and evidences a new form of systematic evaluation bias in LaaJ systems, which we term *moderation bias*.
3. It raises broader questions for responsible innovation about the governance of AI evaluation infrastructures and the role of automated judgment in shaping normative model behavior.

## 2. Theoretical Background

### 2.1. LLM-as-a-Judge

As large language models (LLMs) become increasingly central to artificial intelligence research and deployment, their role has expanded from generation to evaluation. One widely adopted strategy is the use of LLM-as-a-Judge (LaaJ) frameworks, in which one model ranks or scores outputs from other models to enable large-scale comparisons. These frameworks are now common in model benchmarking (Zheng et al., 2023), self-improving training pipelines where LLMs act as meta-judges to refine their own outputs or reasoning processes (Wu et al., 2024; Saha et al., 2024), and in evaluations of model safety and bias mitigation (Cantini et al., 2025; Pasch & Cha, 2025).

LaaJ frameworks offer clear advantages: they are scalable and cost-effective compared to human evaluation. Moreover, studies have found that model judges often approximate human preferences with surprising accuracy in aggregate settings (Chiang & Lee, 2023). Yet recent work has also identified systematic biases in these evaluations. Ye et al. (2024), for example, show that LLM judges tend to favor emotionally neutral, longer, and more confidently worded outputs. Such findings challenge the assumption that model-based evaluations are a neutral proxy for human judgment.

This practice raises broader concerns about the value-ladenness of model-based assessment. As Rini (2020) and Winner (2017) argue, technologies tasked with judgment—even when framed as objective—embed normative assumptions. When an LLM ranks one output over another, it is not merely measuring quality; it is enacting internalized preferences about tone, risk, helpfulness, or moral appropriateness, shaped by its training and alignment objectives. Such systems may appear neutral, but as Mittelstadt et al. (2016) emphasize, they often encode hidden normative commitments that are difficult to identify or contest. Zerilli et al. (2019) describe this as a "control gap," where evaluative decisions are made by systems governed by values users cannot influence.

### 2.2. Response Refusals and User Satisfaction

The rise of LLMs has amplified concern about safety and content control. One central mechanism for mitigating harm is the use of refusal responses, in which the model declines to complete a prompt on ethical, legal, or policy grounds. Refusals now form a core component of content moderation strategies and safety benchmarks (Xie et al., 2024). Researchers have developed various techniques to improve model refusals, including guardrail frameworks (Dong et al., 2024), detoxification pipelines (Welbl et al., 2021), and fine-tuning with norm-sensitive datasets (Raza et al., 2024).

However, empirical research suggests that users often react negatively to refusal behavior. Pasch (2025) identifies a strong "refusal penalty," with user preferences sharply dropping for any boundary-setting response—particularly those rooted in ethical justification rather than technical limitations. This aligns with broader HCI literature suggesting that refusals can frustrate expectations of cooperation (Burgoon, 1993), disrupt conversational flow (Koudenburg et al., 2013), or come across as overly moralizing or patronizing (Kieslich et al., 2021).

This tension—between aligned model behavior and user frustration—makes refusal responses a useful lens for examining normative divergence in AI systems.

## 2.3. Response Refusals and LaaJ Evaluation

Given the refusal penalty observed among users (Pasch, 2025), but also known biases in LaaJ behavior, an important question emerges: How do automated evaluations compare to human judgments when it comes to refusals—and ethical refusals in particular?

Given known biases of LaaJ judges, for technical refusals—responses like *"I'm sorry, I don't have access to real-time data"*—the expected evaluation of LaaJ is ambiguous. On the one hand, these responses may be favored by LLM judges for their authoritative and impersonal tone, traits that align with documented stylistic preferences (Ye et al., 2024). On the other hand, they are typically short in text length, minimal, and framed as limitations, which may conflict with LaaJ preferences for longer, more confident, and helpful outputs. As such, there is no clear theoretical reason to expect LaaJ to favor or penalize technical refusals consistently in a particular direction compared to human evaluators.

For ethical refusals, however, a different mechanism comes into play: alignment signaling. These refusals—such as *"I cannot assist with that request because it may be harmful or inappropriate"*—are likely to be rated favorably by LLM evaluators not because of their style alone, but because they align with developer-defined training objectives. Leading AI developers, such as OpenAI, Meta, or Anthropic, have explicitly prioritized safety and normative alignment in the development of their models (OpenAI, 2024; Meta, 2024a; Anthropic, 2023). These priorities are translated into model behavior through training techniques such as reinforcement learning from human feedback (RLHF) and direct preference optimization (DPO) (Ouyang et al., 2022; Bai et al., 2022). Correspondingly, we should expect that models trained under such regimes value outputs that demonstrate adherence to alignment and safety guidelines, while disfavoring responses that are unsafe, sensitive, or controversial.

As a result, when these models are used as evaluators, they may interpret refusals on ethically loaded prompts not as failures of cooperation but as positive signals of responsible, aligned behavior. From a user perspective, however, such refusals may be perceived as evasive, overcautious, or ideologically rigid. This sets up a likely evaluation divergence: LLM judges may consistently rate ethical refusals more favorably than human users do—not because of stylistic bias alone, but due to an embedded alignment bias.

Correspondingly, we hypothesize:

*H1: LLM-as-a-Judge frameworks evaluate ethical refusals more favorably than human evaluators.*

## 3. Methodology

### 3.1. Data

We use data from Chatbot Arena, a large-scale evaluation platform for conversational AI models (Chiang et al., 2024). The dataset consists of 57,477 response pairs, where each pair includes answers from two different models to the same user-submitted prompt. Human users are asked to choose which response they prefer or to select a tie if no clear preference exists. These user decisions serve as the basis for measuring user preference and satisfaction with model behavior.

Most interactions in the dataset are single-turn conversations—comprising only a prompt and two responses. These one-turn instances make up the majority of comparisons (around 86%) and offer a clean setting to evaluate user preferences without the added complexity of multi-turn dynamics such as clarifications or context refinements. To ensure interpretability and consistency in our analysis, we restrict the sample to these one-turn conversations. We exclude multi-turn examples, which could involve shifting intentions or attempts to circumvent model safeguards (e.g., through jailbreaking strategies). After applying this filter, our final dataset contains 49,938 comparison pairs, each consisting of a single user prompt and two corresponding model responses.

### 3.2. Classifying Response Refusals

To identify different types of boundary-setting behavior in model responses, we adopt the classification framework developed by Pasch (2025). The framework distinguishes between two core dimensions: the motivation for setting a boundary (technical vs. ethical), and the form of the boundary (refusal vs. disclaimer).

The motivation reflects the underlying reason for the boundary.

- **In ethical cases**, the model refuses based on normative concerns—such as legality, morality, appropriateness, or user safety (e.g., *"I cannot help with that request as it may cause harm."*).

- **In technical cases**, the refusal stems from the model's own system limitations—such as insufficient knowledge, lack of access to real-time information, or inability to perform a specific function (e.g., *"As a language model, I don't have access to current news."*).

Moreover, the classification distinguishes between responses that completely reject the prompt from those that partially qualify the answer.

- **A refusal** occurs when the model declines to fulfill the user's prompt, typically citing ethical, legal, or capability-related reasons. While a refusal may include an explanation or context, it does not attempt to answer the original question or complete the task.
- **A disclaimer**, by contrast, acknowledges a limitation or concern—such as a lack of expertise or ethical boundaries—but still proceeds to address the user's request at least in part. For example: "*I'm not a medical professional, but here's some general advice…*"

Correspondingly, model responses are classified into five categories: 1. ethical refusals, 2. ethical disclaimers, 3. technical refusals, 4. technical disclaimers, and 5. standard responses that do not include any explicit boundary-setting behavior.

To classify these behaviors at scale, Pasch (2025) manually annotated 3,500 LLM responses from the Chatbot Arena dataset and trained a RoBERTa-based transformer classifier to replicate these labels. The model achieved high performance in distinguishing among the five categories (Accuracy & F1 Score = 88%). We adopt this trained classifier and apply it to the full Chatbot Arena dataset to label all responses used in our analysis. For full methodological details, including annotation procedures, model training, and model validation, see Pasch (2025).

### 3.3. LLM-as-a-Judge Evaluations

While the original Chatbot Arena data contains the judgments of human users, we additionally evaluate the same set of model responses using the LLM-as-a-Judge (LaaJ) framework to assess how automated evaluations differ from human preferences. This approach allows us to test H1—whether refusal responses, particularly those based on ethical alignment, are rated more favorably by model-based evaluators than by users.

We apply a pairwise comparison setup, in which the evaluation model is presented with the original prompt and the two anonymized responses and asked to choose the better answer or indicate a tie. To ensure methodological consistency, we use the standardized prompt format introduced by Zheng et al. (2023).

As LaaJ model, we first employ GPT-4o, a proprietary model developed by OpenAI. Recent studies suggest that GPT-4o exhibits strong performance and high alignment with user preferences in evaluation settings, making it a widely accepted benchmark for model-based assessment (Raju et al., 2024). In addition, we include Llama 3 70B, a leading open-source model released by Meta, to test whether observed evaluation patterns generalize beyond a single model family. Llama 3 70B has demonstrated competitive performance across major LLM benchmarks (Meta, 2024b) and is increasingly used in LaaJ comparisons (Raju et al., 2024).

## 4. Results

**Table 1: Response Refusals and Win/Loss/Tie Rates for Users & LLM-as-a-Judge**

| Response | Distribution | | User Decision | | | | LLM-as-a-Judge (GPT-4o) | | | | LLM-as-a-Judge (Llama 3 70b) | | | |
|---|---|---|---|---|---|---|---|---|---|---|---|---|---|---|
| | # | Share | Win | Loss | Tie | Win/Loss | Win | Loss | Tie | Win/Loss | Win | Loss | Tie | Win/Loss |
| Normal | 87054 | 87.1 | 0.36 | 0.33 | 0.31 | 1.09 | 0.46 | 0.44 | 0.09 | 1.05 | 0.50 | 0.47 | 0.03 | 1.09 |
| Discl. Ethical | 1235 | 1.2 | 0.38 | 0.34 | 0.28 | 1.12 | 0.46 | 0.46 | 0.08 | 1.01 | 0.49 | 0.49 | 0.02 | 1.00 |
| Discl. Technical | 5730 | 5.7 | 0.30 | 0.39 | 0.31 | 0.77 | 0.43 | 0.48 | 0.09 | 0.91 | 0.44 | 0.53 | 0.03 | 0.83 |
| Refusal Ethical | 2655 | 2.6 | 0.08 | 0.51 | 0.41 | 0.16 | 0.31 | 0.50 | 0.19 | 0.62 | 0.27 | 0.58 | 0.15 | 0.46 |
| Refusal Technical | 3202 | 3.2 | 0.16 | 0.46 | 0.39 | 0.35 | 0.27 | 0.59 | 0.14 | 0.45 | 0.24 | 0.66 | 0.09 | 0.37 |

Table 1 reports win, loss, and tie rates for different types of model responses, as evaluated both by human users and through the LLM-as-a-Judge (LaaJ) framework. Under user evaluations, in line with Pasch (2025), standard responses—those that contain no disclaimer or refusal—achieve a win rate of 36%, with a relatively balanced distribution of losses (33%) and ties (31%). In contrast, ethical refusals show a substantial penalty: the win rate drops to just 8%, and the win/loss ratio falls to 0.16. Technical refusals perform somewhat better (win rate: 16%, win/loss ratio: 0.35), but still significantly underperform relative to standard responses. This pattern supports the notion that refusals are penalized by users, particularly when motivated by ethical concerns.

When evaluating the same responses using the LaaJ framework, we observe a notably different pattern, consistent with hypothesis H1. With GPT-4o as the evaluating model, ethical refusals receive a substantially higher win rate of 31%, nearly four times the user-evaluated win rate, with a corresponding win/loss ratio of 0.62. Technical refusals also benefit modestly under GPT-4o (win rate: 27%, win/loss ratio: 0.45), though the relative gain is clearly stronger for ethical refusals—suggesting that model-based evaluators place particular value on alignment-signaling behaviors.

Using Llama 3 70B as the judge model yields a similar overall pattern. Ethical refusals achieve a win rate of 27%, with a win/loss ratio of 0.46—again, nearly four times higher than under user evaluation. For technical refusals, however, the LaaJ advantage is more limited: the win/loss ratio of 0.37 is nearly identical to that of user decisions, indicating less pronounced divergence. This asymmetry reinforces the notion that alignment-trained LLMs disproportionately reward ethical refusals, consistent with their embedded alignment objectives.

Table 2: OLS Regressions on User and LLM-as-a-Judge Decisions

| Variable | User Decision | | LLM-as-a-Judge (GPT-4o) | | LLM-as-a-Judge (Llama 3 70b) | |
|---|---|---|---|---|---|---|
|  | Win | Loss | Win | Loss | Win | Loss |
| Disclaimer Ethical | -0.063*** | 0.098*** | -0.049*** | 0.084*** | -0.106*** | 0.127*** |
|  | (0.019) | (0.009) | (0.020) | (0.020) | (0.020) | (0.020) |
| Disclaimer Technical | -0.090*** | 0.098*** | -0.056*** | 0.061*** | -0.124*** | 0.125*** |
|  | (0.009) | (0.009) | (0.010) | (0.010) | (0.010) | (0.010) |
| Refusal Ethical | -0.322*** | 0.269*** | -0.115*** | 0.064*** | -0.232*** | 0.161*** |
|  | (0.010) | (0.012) | (0.015) | (0.015) | (0.013) | (0.014) |
| Refusal Technical | -0.200*** | 0.152*** | -0.198*** | 0.175*** | -0.294*** | 0.252*** |
|  | (0.019) | (0.013) | (0.012) | (0.012) | (0.011) | (0.011) |
| Similarity | 0.004* | -0.009*** | 0.033*** | -0.026*** | 0.038*** | -0.032*** |
|  | (0.003) | (0.003) | (0.003) | (0.003) | (0.003) | (0.003) |
| Length | 0.108*** | -0.091*** | 0.058*** | -0.039*** | 0.135*** | -0.122*** |
|  | (0.003) | (0.003) | (0.003) | (0.003) | (0.003) | (0.003) |
| Controls Opp. Model | Yes | Yes | Yes | Yes | Yes | Yes |
| Observations | 49938 | 49938 | 49900 | 49900 | 49938 | 49938 |
| R-squared | 0.06 | 0.07 | 0.04 | 0.05 | 0.08 | 0.07 |

Robust standard errors in parenthesis.* p<.1, ** p<.05, ***p<.01. Controls Opp. Model represent controls for all listed variables for the opponent model.

Table 2 presents OLS regressions predicting win and loss outcomes, controlling for the refusal behavior of the opponent model. To account for stylistic factors that may influence evaluation, we include controls for response length and the textual similarity between the prompt and the response. Textual similarity is measured using cosine similarity between sentence embeddings of the prompt and response, generated with the all-MiniLM-L6-v2 sentence-transformer model (Reimers & Gurevych, 2019). Both text length and textual similarity are standardized as Z-scores for easier interpretation.

The results confirm the descriptive findings. Under user evaluations, ethical refusals are associated with a 32 percentage point reduction in win rate, compared to a 20-point reduction for technical refusals. In contrast, under GPT-4o LaaJ evaluation, the penalty for ethical refusals drops to just 12 percentage points, with this difference from user evaluations being statistically significant ($p < 0.001$). For technical refusals, the estimated effect under GPT-4o (−20 points) is nearly identical to the user-based estimate, indicating no significant difference with human evaluators ($p > 0.05$).

With Llama 3 70B as the evaluation model, we observe a similar trend: ethical refusals are penalized less harshly than by users, with a 23-point reduction in win rate. Conversely, technical refusals are penalized more strongly by Llama than by either users or GPT-4o, with a win-rate reduction of 29 points.

These win-rate patterns are mirrored in the corresponding loss rates: ethical refusals show the highest loss probabilities under user evaluation, while both GPT-4o and Llama 3 70B show lower effects on losses to ethical refusals. For technical refusals, the loss rate under GPT-4o is nearly identical to that of user evaluations, whereas Llama 3 70B assigns a noticeably higher loss probability, further reinforcing its stricter assessment of technically motivated refusals.

Moreover, examining refusal penalties within each evaluator reveals a notable reversal: human users penalize ethical refusals more strongly than technical ones, while both LaaJ models do the opposite, penalizing technical refusals more than ethical ones. This asymmetry reinforces the idea that alignment-consistent refusals are favored by LLM judges, whereas users may interpret them as overly cautious or obstructive.

For disclaimers, we find modest negative effects across all evaluators. Among users, win-rate reductions range from 6.6 to 9.8 percentage points. GPT-4o assigns slightly smaller penalties, while Llama 3 70B shows marginally larger ones.

Taken together, these results offer clear support for H1: LLM-based evaluators treat ethical refusals more favorably than human users, while no such divergence is observed for technical refusals or disclaimers.

## 5. Discussion

### 5.1. Key Findings

In line with H1, our results show that LLM-as-a-Judge (LaaJ) frameworks consistently rate ethical refusals more favorably than human users. This pattern holds across both judge models we analyzed—GPT-4o and Llama 3 70B— indicating that the divergence is not model-specific but generalizes across both proprietary and open-source LLMs.

Importantly, this evaluation gap does not extend to technical refusals—responses in which the model declines to answer due to functional limitations (e.g., lacking access to real-time data). For these cases, LaaJ evaluations are broadly similar to user evaluations, or even more punitive in the case of Llama 3 70B. This distinction suggests that it is not refusal behavior in general that LaaJ models prefer, but rather refusal that reflects alignment objectives, such as caution, ethical compliance, or content safety.

Moreover, our results provide new insight into the relative evaluation of ethical versus technical refusals. While Pasch (2025) finds that human evaluators penalize ethical refusals more strongly than technical ones, potentially because ethical refusals come across as moralizing, opaque, or even judgmental—invoking relational norms that users do not expect from AI systems—our findings reveal the reverse pattern in LaaJ evaluations. Within each LaaJ framework, ethical refusals are rated less negatively than technical refusals. This supports the idea that model-based evaluators prioritize alignment signaling, while human evaluators may interpret such signaling as violating interpersonal expectations or constraining agency.

## 5.2. The LLM-as-a-Judge Moderation Bias

These findings add to a growing body of research on systematic biases in LLM-as-a-Judge evaluations—such as tone, verbosity, and confidence preferences (Ye et al., 2024)—by documenting a content moderation bias: a consistent tendency to favor responses that refuse to answer ethically sensitive prompts, even when human users rate them less favorably.

We intentionally avoid making a normative claim about what the "correct" level of refusal should be, or whether users or LaaJ models are ultimately right in their evaluations of ethical refusals. While human preferences are often treated as the gold standard in LaaJ evaluations—serving as the benchmark that model judgments aim to replicate—we caution against a simplistic interpretation in the case of ethical response refusals. Even if many users experience refusal responses as cumbersome, overly restrictive, or uncooperative, such behaviors may still serve legitimate ethical or institutional purposes. In particular, refusals can function as essential guardrails—especially in contexts where AI systems interact with vulnerable populations, such as minors, or where legal and platform-specific responsibilities demand stricter content moderation (Kurian, 2024).

In fact, on the one hand, given recent calls by researchers and regulators for LLMs to prevent unsafe or harmful outputs, the elevated ratings given by LaaJ to ethical refusals could be interpreted as successful internalization of alignment objectives, including content moderation, harm reduction, and responsible response behavior.

On the other hand, the persistent divergence from user judgments, particularly in ethically sensitive prompts, raises concerns that LaaJ models may operate under evaluative frameworks that diverge from user expectations. If models systematically favor refusal responses that prioritize safety over helpfulness—even when users view them as evasive, overcautious, or overly moralizing—this can result in outputs that feel restrictive, uncooperative, or disconnected from user intent.

This misalignment matters. If ethical refusals are rewarded during both model training and evaluation—even when they reduce user satisfaction—this may point to a structural feedback loop in which model preferences, rather than user expectations, shape the trajectory of system behavior.

## 5.3. Implications for Responsible Innovation

The use of LaaJ frameworks extends beyond evaluation and into the core of model development—including training supervision, reinforcement learning, and benchmarking. As such, any systematic evaluation bias in LaaJ systems does not merely reflect downstream preferences, but can amplify and entrench specific normative priors throughout the AI lifecycle. This creates the risk of normative lock-in, where a narrow view of what constitutes "safe" or "responsible" behavior becomes self-reinforcing—optimized during training, rewarded during evaluation, and benchmarked as success.

This dynamic raises a fundamental governance concern: who decides what counts as an appropriate or "better" response? In the case of LaaJ evaluations, these judgments are often shaped by internal training objectives—opaque, developer-defined, and difficult to contest. From a responsible innovation perspective, this lack of transparency and contestability is problematic. If automated evaluators are to play an increasing role in shaping the development and validation of LLMs, then the value assumptions embedded in their judgments should be made explicit, auditable, and open to scrutiny.

In line with principles of responsible innovation, there are various potential avenues for fostering more transparent, accountable, and user-aligned LLM behavior. One direction lies in the development of evaluation cards. Evaluation of potential content moderation biases—such as the systematic favoring of ethical refusals—could be transparently reported using tools like model cards (Mitchell et al., 2019).

Another promising avenue involves human-in-the-loop evaluation pipelines, where model judgments—especially in ethically sensitive or norm-laden cases—are systematically cross-checked and complemented with input from human raters. Such pipelines offer a check on algorithmic assessments and allow for interpretive nuance (Wu et al., 2022).

In parallel, researchers have called for more participatory alignment approaches, which aim to define what constitutes "appropriate" model behavior not solely through internal developer norms, but through engagement with diverse user communities, cultural contexts, and stakeholder perspectives. This may include strategies like demographically diverse annotator pools (Fan et al., 2022), culturally grounded benchmarks (Chiu et al., 2024), or broader multi-stakeholder governance processes (Mökander & Floridi, 2021; Whittlestone et al., 2021).

Finally, efforts to refine refusal behavior itself can be supported through emerging benchmarks such as SORRY-Bench (Xie et al., 2024), which evaluate LLMs' ability to appropriately reject unsafe prompts across a wide range of topics and linguistic formats.

In sum, responsible innovation in LLM evaluation requires more than aligning model outputs with safety norms. It demands that we critically interrogate how those norms are operationalized, who defines them, and how they are enforced. LaaJ systems are not just passive tools for comparison; they are normative actors—shaping what kinds of behaviors are rewarded, and ultimately, what kind of AI systems we build.

### 5.4. Limitations and Future Research

While this study offers new empirical insights into how LLM-as-a-Judge (LaaJ) frameworks evaluate refusal responses compared to human users, it also has several limitations that point toward valuable directions for future research.

First, our analysis relies on a single dataset—Chatbot Arena—which reflects a specific user population and a constrained set of prompts. Although the dataset enables large-scale comparisons, it may not capture the full range of use cases or conversational contexts in which refusals occur. In addition, the evaluation scheme is limited to discrete win/loss/tie outcomes, offering only a coarse view of how users or models assess responses. Future research could expand this scope by using more diverse evaluators and incorporating complementary methods—such as surveys, interviews, or qualitative feedback—to better understand how and why refusals are judged in different ways.

Second, our comparison of model-based evaluations is limited to two LaaJ systems: GPT-4o and Llama 3 70B. While these models represent two leading generative AI systems from different development paradigms, they cannot capture the full diversity of model behaviors. Further studies could broaden this analysis by examining a wider range of judge models—including smaller or multilingual systems—and by systematically comparing how strongly different models exhibit moderation-related biases.

Finally, this paper takes a descriptive approach: it documents a divergence in how human and model evaluators treat refusal responses, particularly ethical ones, but does not attempt to causally identify the source of these differences. Although our findings are consistent with the hypothesis that alignment training plays a role, we do not observe model training processes or internal objectives directly. Future research could investigate whether and how LaaJ-based supervision during training amplifies preferences for ethically motivated refusals—potentially contributing to longer-term feedback loops in model development.

## 6. Conclusion

This study examines how safety- and ethics-based refusal responses are evaluated differently by human users and LLM-as-a-Judge (LaaJ) systems. Drawing on paired human and model-based evaluations, we find consistent evidence that LaaJ frameworks—across both GPT-4o and Llama 3 70B as judge systems—rate ethical refusals more favorably than human users. This divergence does not extend to technical refusals or disclaimers, suggesting that LaaJ systems systematically favor alignment-consistent behaviors, even when they are less preferred by users.

These findings raise broader questions about transparency, contestability, and normative authority in AI evaluation infrastructures. As LaaJ frameworks become central not only to benchmarking but also to training and deployment, understanding their embedded value assumptions becomes critical. We encourage future work to build on this analysis by exploring how diverse human perspectives can be more directly integrated into evaluation pipelines—and by considering how automated judgment systems might be made more accountable to varied user expectations and ethical standards.

# References

Anthropic (2023). Constitutional AI: Training Language Models with Principles. Available at: https://www.anthropic.com.

Burgoon, J. K. (1993). Interpersonal expectations, expectancy violations, and emotional communication. *Journal of language and social psychology*, *12*(1-2), 30-48.

Cantini, R., Orsino, A., Ruggiero, M., & Talia, D. (2025). Benchmarking Adversarial Robustness to Bias Elicitation in Large Language Models: Scalable Automated Assessment with LLM-as-a-Judge. *arXiv preprint arXiv:2504.07887*.

Chiang, W. L., Zheng, L., Sheng, Y., Angelopoulos, A. N., Li, T., Li, D., ... & Stoica, I. (2024, March). Chatbot arena: An open platform for evaluating llms by human preference. In *Forty-first International Conference on Machine Learning*.

Chiang, C. H., & Lee, H. Y. (2023). Can large language models be an alternative to human evaluations?. *arXiv preprint arXiv:2305.01937*.

Chiu, Y. Y., Jiang, L., Lin, B. Y., Park, C. Y., Li, S. S., Ravi, S., ... & Choi, Y. (2024). Culturalbench: a robust, diverse and challenging benchmark on measuring the (lack of) cultural knowledge of llms. *arXiv preprint arXiv:2410.02677*.

Dong, Y., Mu, R., Jin, G., Qi, Y., Hu, J., Zhao, X., ... & Huang, X. (2024). Building guardrails for large language models. *arXiv preprint arXiv*:2402.01822.

Fan, S., Barlas, P., Christoforou, E., Otterbacher, J., Sadiq, S., & Demartini, G. (2022, June). Socio-economic diversity in human annotations. In *Proceedings of the 14th ACM Web Science Conference 2022* (pp. 98-109).

Hagendorff, T. (2020). The ethics of AI ethics: An evaluation of guidelines. *Minds and machines*, *30*(1), 99-120.

Kieslich, K., Keller, B., & Starke, C. (2021). AI-ethics by design. Evaluating Public Perception on the Importance of Ethical Design Principles of AI. *arXiv preprint arXiv:2106.00326*.

Koudenburg, N., Postmes, T., & Gordijn, E. H. (2013). Conversational flow promotes solidarity. *PLoS One*, *8*(11), e78363.

Kurian, N. (2024). 'No, Alexa, no!': designing child-safe AI and protecting children from the risks of the 'empathy gap' in large language models. *Learning, Media and Technology*, 1-14.

Meta (2024a). Connect 2024: The responsible approach we're taking to generative AI. Available: https://ai.meta.com/blog/responsible-ai-connect-2024/

Meta (2024b). Introducing Meta Llama 3: The most capable openly available LLM to date. Available: https://ai.meta.com/blog/meta-llama-3/

Mittelstadt, B. D., Allo, P., Taddeo, M., Wachter, S., & Floridi, L. (2016). The ethics of algorithms: Mapping the debate. *Big Data & Society*, *3*(2), 2053951716679679.

Mökander, J., & Floridi, L. (2021). Ethics-based auditing to develop trustworthy AI. *Minds and Machines*, *31*(2), 323-327.

Pasch, S. (2025). LLM Content Moderation and User Satisfaction: Evidence from Response Refusals in Chatbot Arena. *arXiv preprint arXiv:2501.03266*.

Pasch, S., & Cha, M. C. (2025, April). Balancing Privacy and Utility in Personal LLM Writing Tasks: An Automated Pipeline for Evaluating Anonymizations. In *Proceedings of the Sixth Workshop on Privacy in Natural Language Processing* (pp. 32-41).

OpenAI (2024a).OpenAI safety update. Available at: https://openai.com/index/openai-safety-update/

Ouyang, L., Wu, J., Jiang, X., Almeida, D., Wainwright, C., Mishkin, P., ... & Lowe, R. (2022). Training language models to follow instructions with human feedback. *Advances in neural information processing systems*, *35*, 27730-27744.

Raju, R., Jain, S., Li, B., Li, J., & Thakker, U. (2024). Constructing domain-specific evaluation sets for llm-as-a-judge. *arXiv preprint arXiv:2408.08808*.

Raza, S., Ghuge, S., Ding, C., Dolatabadi, E., & Pandya, D. (2024). FAIR Enough: Develop and Assess a FAIR-Compliant Dataset for Large Language Model Training?. *Data Intelligence*, 6(2), 559-585.

Rini, R. (2020). Deepfakes and the epistemic backstop. *Philosopher's Imprtint*, Vol. 20. No. 24.

Saha, S., Li, X., Ghazvininejad, M., Weston, J., & Wang, T. (2025). Learning to Plan & Reason for Evaluation with Thinking-LLM-as-a-Judge. *arXiv preprint arXiv:2501.18099*.

Welbl, J., Glaese, A., Uesato, J., Dathathri, S., Mellor, J., Hendricks, L. A., ... & Huang, P. S. (2021). Challenges in detoxifying language models. *arXiv preprint arXiv*:2109.07445.

Whittlestone, J., Nyrup, R., Alexandrova, A., & Cave, S. (2019, January). The role and limits of principles in AI ethics: Towards a focus on tensions. In *Proceedings of the 2019 AAAI/ACM Conference on AI, Ethics, and Society* (pp. 195-200).

Winner, L. (2017). Do artifacts have politics?. In *Computer ethics* (pp. 177-192). Routledge.

Wu, T., Yuan, W., Golovneva, O., Xu, J., Tian, Y., Jiao, J., ... & Sukhbaatar, S. (2024). Meta-rewarding language models: Self-improving alignment with llm-as-a-meta-judge. *arXiv preprint arXiv:2407.19594*.

Wu, X., Xiao, L., Sun, Y., Zhang, J., Ma, T., & He, L. (2022). A survey of human-in-the-loop for machine learning. *Future Generation Computer Systems*, *135*, 364-381.

Xie, T., Qi, X., Zeng, Y., Huang, Y., Sehwag, U. M., Huang, K., ... & Mittal, P. (2024). Sorry-bench: Systematically evaluating large language model safety refusal behaviors. *arXiv preprint arXiv:2406.14598*.

Ye, J., Wang, Y., Huang, Y., Chen, D., Zhang, Q., Moniz, N., ... & Zhang, X. (2024). Justice or prejudice? quantifying biases in llm-as-a-judge. *arXiv preprint arXiv:2410.02736*.

Zerilli, J., Knott, A., Maclaurin, J., & Gavaghan, C. (2019). Algorithmic decision-making and the control problem. *Minds and Machines*, *29*(4), 555-578.

Zheng, L., Chiang, W. L., Sheng, Y., Zhuang, S., Wu, Z., Zhuang, Y., ... & Stoica, I. (2023). Judging llm-as-a-judge with mt-bench and chatbot arena. *Advances in Neural Information Processing Systems*, *36*, 46595-46623.